\documentclass{llncs}
\usepackage[utf8]{inputenc}
\usepackage{graphicx}
\usepackage{moreverb}
\usepackage{rotating}

\def\Tabvrule{\vrule width-0.4pt}               
\def\Tabhrule{\hrule \hrule height-0.4pt}       
\def\Tabstrut{\vrule height2.2ex                
       depth0.8ex                               
       width0ex                                 
\relax}

\def\PasCase#1{\omit%
$\vcenter{\hbox{\vbox to 0.4pt{}}\hbox{\makebox[3ex]{\Tabstrut$#1$}}}%
\Tabvrule$}
\def\PasCasePoint{\PasCase{\cdot}}
\def\DessinCarre#1{\vcenter{\hbox{}\hrule%
\hbox{\vrule\makebox[3ex]{\Tabstrut$#1$}\vrule}\Tabhrule}%
\Tabvrule}
\def\GenRuban#1{\vcenter{\halign{&$\DessinCarre{##}$\cr#1}}\egroup}

\def\sTabvrule{\vrule width-0.4pt}
\def\sTabhrule{\hrule \hrule height-0.4pt}
\def\sTabstrut{\vrule height1.6ex depth0.6ex width0ex \relax}
\def\sDessinCarre#1{\vcenter{\hbox{}\hrule%
\hbox{\vrule\makebox[2.3ex]{\sTabstrut$\scriptstyle#1$}\vrule}\sTabhrule}%
\sTabvrule}
\def\sGenRuban#1{\vcenter{\halign{&$\sDessinCarre{##}$\cr#1}}\egroup}

\def\ruban{%
\bgroup\let\ =\omit
\let\\=\cr
\let\.=\PasCasePoint
\offinterlineskip
\GenRuban}

\def\sruban{%
\bgroup\let\ =\omit
\let\\=\cr
\let\.=\PasCasePoint
\offinterlineskip
\sGenRuban}
\begin{document}
\pagestyle{plain}
\title{Formal Proof of SCHUR Conjugate Function%
\thanks{This research is supported by the PEPS-CNRS project CerBISS.}
\thanks{The final publication of this paper is
   available at www.springerlink.com}
}
\author{Franck Butelle\inst{1} \and Florent Hivert\inst{2} \and
  Micaela Mayero\inst{1,}\inst{3} \and Frédéric Toumazet\inst{4}}
\authorrunning{F. Butelle, F. Hivert, M. Mayero, F. Toumazet}
\institute{LIPN UMR 7030, Université Paris 13, 
  Villetaneuse, F-93430
  \and LITIS EA 4108, Université de Rouen, Saint-Etienne-du-Rouvray, F-76801
\and LIP, INRIA Grenoble -- Rhône-Alpes, UMR 5668,
  UCBL, ENS Lyon, Lyon, F-69364 
  \and LIGM UMR 8049, Université de Marne-la-Vallée, Champs sur Marne,
  F-77454  
}
\maketitle

\begin{abstract}
The main goal of our work is to formally prove the correctness of the key
commands of the SCHUR software, an interactive program for calculating
with characters of Lie groups and symmetric functions. The core of the
computations relies on enumeration and manipulation of combinatorial
structures. As a first "proof of concept", we present a formal proof of the
conjugate function, written in C. This function computes the conjugate
of an integer partition. To formally prove this program, we use the Frama-C
software. It allows us to annotate C functions and to generate proof
obligations, which are proved using several automated theorem provers.
In this paper, we also draw on methodology, discussing on how to 
formally prove this kind of program.
\end{abstract}

\section{Introduction}

SCHUR~\cite{SCHUR} is an interactive software for calculating properties of
Lie groups and symmetric functions~\cite{MacDonald79}.  It is used in research
by combinatorists, physicists, theoretical chemists~\cite{bookSCHUR} 
as well as for educational
purpose as a learning tool for students in algebraic combinatorics. One of
its main uses is to state conjectures on combinatorial objects.
For such use, it is important to have some confidence
in the results produced by SCHUR.

Until now, the method used to get some confidence in the
results has mostly been based on just one example for each command.

The computation of other examples is complex due to the well known
combinatorial explosion, especially when using algorithms associated to the
symmetric group, see section~\ref{section:SCHUR}. Unfortunately, the
combinatorial explosion as well as computing time forbid test generation or
verification techniques (model checking).  Therefore, in this paper, we focus on formal proof
of the existing program.

With the aim of verifying the whole software, we start with proving
the correctness of its fundamentals bricks.
The main combinatorial object used in SCHUR is integer partition.
The first non-trivial operation on integer partitions is
the conjugate. Moreover, conjugate function is necessary
for more than half of the 240 interactive commands of SCHUR.
From this point of view, we can say that conjugate is
a critical function of SCHUR.

The very first work consists in isolating (see~\ref{section:Difficulties})
the code of this function from the program. Next, 
we chose to use the most popular tool of program proof community's
Frama-C~\cite{Frama-C} successor of Caduceus.
Frama-C is a plug-in system. In order to
prove programs, we used Jessie~\cite{Jessie}, the deductive
verification plug-in of~C programs annotated with ACSL~\cite{ACSL}. 
The generated verification conditions can be
submitted to external automatic provers,
and for more complex situations, to interactive theorem provers
as well (see section~\ref{section:Frama-C}).

After a short presentation of software tools and theoretical concepts,
we will present the formal proof of a program. 
Finally, after discussing difficulties and mistakes encountered along the way, we
will propose a methodology to prove such a software, and finally discuss 
future work. 

\section{Presentation of the Software Used}
\subsection{The SCHUR Software}\label{section:SCHUR}

SCHUR is an interactive software 
for calculating properties of Lie groups and symmetric functions.
A Symmetric Function is a function which is symmetric, or invariant, 
under any permutation of its variables. For example 
$f(x_1,x_2,x_3)= x_1+x_2+x_3$ is a symmetric function.

SCHUR has originally written by Prof. Brian G. Wybourne in Pascal
language. Then it was translated into C by an automatic program
making it quite difficult to read. There are almost no
comments in the code, the code is more than 50,000 lines long with many
global variables. Local variables have names such as $W52$ and so on.

After the death of Prof. Wybourne in November 2003, some people felt
that his program should be maintained, and if possible enhanced,
with a view to making it freely available to the mathematics and physics
research community.

Nowadays, it is open source under the GPL license and includes
more than 240 commands. The code still includes very few
comments. Some mistakes have been corrected but some
interactive commands are so intricate that it is difficult to
have more than a few examples to check them against and most people
do not even know if the result is correct or not.

This is why we started to work on this code. Firstly
some of the commands in SCHUR are very well implemented
(for example, plethysm is computed faster by SCHUR than by many other
combinatorial toolboxes). Formally proving some key functions inside
would also be a major advance for its research community.

\subsection{The Frama-C Software}\label{section:Frama-C}

Frama-C \cite{Frama-C} is an open source extensible platform dedicated to
source code analysis of~C software. It is co-developed by two French
public institutions: CEA–LIST (Software Reliability Laboratory) and
INRIA-Saclay (ProVal project). 

Frama-C is a plug-in system. In order to 
prove programs, we use Jessie \cite{Jessie}, the deductive
verification plug-in of~C programs annotated with ACSL \cite{ACSL}. It
uses internally the languages and tools of the Why
platform \cite{Why}. The Jessie plug-in uses Hoare-style~\cite{Hoare69}
weakest
precondition computations to formally prove ACSL properties.
The generated verification conditions (VC) can be
submitted to external automatic provers such as
Simplify \cite{Simplify}, Alt-Ergo \cite{AltErgo}, Z3 \cite{Z3},
CVC3 \cite{CVC3}.

These automatic provers belong to SMT (Satisfiability
Modulo Theories) solvers. The SMT problem is a decision problem for
logical formulas with respect to combinations of background theories
expressed in classical first-order logic with equality. First-order
logic is undecidable. Due to this high computational difficulty, it is
not possible to build a procedure that can solve arbitrary SMT
problems. Therefore, most procedures focus on the more realistic goal
of efficiently solving problems that occur in practice.

For more complex situations, interactive theorem provers can be used
to establish the validity of VCs, like Coq \cite{Coq}, PVS \cite{PVS},
Isabelle/HOL \cite{IsabelleHOL}, etc. For our purpose, we used Coq (see 
section~\ref{section:proof}) since it is the one best known to the authors.

\section{The Conjugate Function}

In this section, the basics of algebraic combinatorics are given so 
that the reader can understand what is actually proved. Interestingly in this
field, though the interpretation of what is actually computed can be of a very
abstract algebraic level, the computation itself boils down most of the time
to possibly intricate but rather elementary manipulations. 

\subsection{Combinatorial and Algebraic Background: Integer Partitions}
A \textbf{partition} of a positive integer $n$ is a way of writing 
$n$ as a sum of a non-increasing sequence of integers. 
For example $\lambda=(4,2,2,1)$ and $\mu=(2,1)$ are partitions 
of $n=9$ and $n'=3$ respectively. 
We write $|\lambda|=n$ and $|\mu|=n'$~\cite{Andrews84}. 

The \textbf{Ferrers diagram} $F^{\lambda}$ associated to a partition 
$\lambda=(\lambda_1,\lambda_2,...,\lambda_p)$ consists of $|\lambda|=n$ boxes, 
arranged in $l(\lambda)=p$ left-justified rows of lengths 
$\lambda_1,~\lambda_2,~...,~\lambda_p$. 
Rows in $F^{\lambda}$ are oriented downwards (or upwards for some authors).
$F^{\lambda}$ is called the shape of $\lambda$. 

\begin{definition}
  The conjugate of an integer partition is the partition associated to the
  diagonal symmetric of its shape.
\end{definition}
For example, for $\lambda=(3,2,1,1,1)$, here is
the Ferrers diagram $F^{\lambda}$
and the Ferrers diagram of the conjugate partition:

$$ \ruban{ & &\\ & \\ \\\\\\} \qquad
\ruban{ & & & &\\ & \\\\}$$

So the conjugate partition of $(3,2,1,1,1)$ is $(5,2,1)$.

A \textbf{semi-standard Young tableau} of shape $\lambda$ 
is a numbering of the boxes of $F^{\lambda}$ with entries from $\{1,2,...,n\}$, 
weakly increasing across rows and strictly increasing down columns. 
A tableau is \textbf{standard} if and only if each entry appears
only once. Here is an example of shape $(4,2,2,1)$ tableau:
$$
\ruban{1 & 2 & 2 & 5\\2 & 4\\3 & 6\\5\\}
$$

\sloppypar
A \textbf{symmetric function} of a set of variables $\{x_1, x_2,
\dots\}$ is a function $f(x_1,x_2,\dots)$ of those variables which is
invariant under any permutation of those variables (that is for example
$f(x_1,x_2,\dots) = f(x_2,x_1,\dots)$). This definition is usually restricted to polynomial
functions. The most important linear basis of symmetric
function's algebra is called the \textbf{Schur functions} and they are combinatorially
defined as follows: for a given semi-standard Young tableau $T$ of shape
$\lambda$, write ${\mathbf x}^T$ the product of the $x_i$ for all $i$ appearing in the
tableau. Then

\begin{equation}
   s_\lambda({\mathbf x}) = \sum_{T\in \mathrm{Tab}(\lambda)}{{\mathbf x}^T}.
\end{equation}
where $\mathrm{Tab}(\lambda)$ is the set of all tableaux of shape
$\lambda$. We will note $s_\lambda({\mathbf x})$, $s_\lambda$.
For example, consider 
the tableaux of shape $(2,1)$ using just three variables ${x_1,x_2,x_3}$:
$$
\ruban{1&1 \\ 2 \\} \quad
\ruban{1&1 \\ 3 \\} \quad
\ruban{2&2 \\ 3 \\} \quad
\ruban{1&2 \\ 3 \\} \quad
\ruban{1&3 \\ 2 \\} \quad
\ruban{1&2 \\ 2 \\} \quad
\ruban{1&3 \\ 3 \\} \quad
\ruban{2&3 \\ 3 \\}
$$
The associated Schur function is therefore:
\begin{eqnarray}\label{eq:s21x1x2x3}
s_{(21)}(x_1,x_2,x_3) = x_1^2x_2 + x_1^2x_3 + x_2^2x_3 + 2x_1x_2x_3 +
x_1x_2^2 + x_1x_3^2 + x_2x_3^2
\end{eqnarray}
thus:
$$
s_{(21)} = s_{(21)}(x_1,x_2,x_3) + s_{(21)}(x_1,x_2,x_3,x_4) +\dots
$$
Note that, with this combinatorial definition, 
the symmetry of $s_{(21)}(x_1,x_2,x_3)$  is not exactly obvious.

We need to recall some well-known results in symmetric function theory:
though Schur functions have historically been defined by
Jacobi~\cite{Jacobi41}, they were named in the honor of Schur who
discovered their crucial role in the representation theory of the
symmetric group and the general linear group. Namely, after the discovery by
Frobenius that the irreducible representation of the symmetric groups
are indexed by integer partitions, Schur showed that those functions
can be interpreted as characters of those irreducible representation,
and by Schur-Weyl duality characters of Lie groups and Lie algebras. 
Notably we obtain the representation of
the general linear groups ($GL_n$) and unitary groups
($U_n$)~\cite{Littlewood50} from the symmetric group representations.
In this setting, the conjugate of the partition essentially
encodes the tensor product of a representation by the sign representation.

Further work by Schur-Littlewood involve infinite sum of Schur functions
associated to partitions~\cite{LascouxPragacz88}, 
whose conjugates have a particular form. In
particular, these series are used to obtain symplectic ($Sp_{2n}$) and
orthogonal character groups ($O_n$) (symmetric and orthogonal Schur
functions) from standard Schur functions~\cite{Newell51}.

One particularly important and difficult computational problem here 
is plethysm (see SCHUR reference manual~\cite{SCHUR} and~\cite{MacDonald79}). 
It is the analogue in symmetric functions 
of the substitution of
polynomial inside another polynomial $f(x)\mapsto f(g(x))$. It is called
plethysm because by some combinatorial explosion, 
it involves very quickly a
lot (a plethora) of terms, making it something very difficult to compute
efficiently. For example, $s_{(21)}(s_{(21)})$, the first example with non trivial partitions in the input 
is already very hard to compute by hand. 
First we can regard $s_{(21)}$ as a function in as many monomial as
in (\ref{eq:s21x1x2x3}):
\begin{eqnarray*}
s_{(21)}(s_{(21)})(x_1,x_2,x_3) &=& 
s_{(21)}(x_1^2x_2, x_1^2x_3, x_2^2x_3, x_1x_2x_3, x_1x_2x_3,
x_1x_2^2, x_1x_3^2, x_2x_3^2)
\end{eqnarray*}
it can be shown that the following holds:
\begin{eqnarray*}
s_{(21)}(s_{(21)}) &=& s_{(22221)} + s_{(321111)} + 2s_{(32211)} + s_{(3222)} + s_{(33111)} + \\
&&3s_{(3321)} + 2s_{(42111)} + 3s_{(4221)} + 3s_{(4311)} +
3s_{(432)} +\\
&& s_{(441)} + s_{(51111)} + 2s_{(5211)} + s_{(522)} + 2s_{(531)} + s_{(54)} + s_{(621)}
\end{eqnarray*}

\subsection{Computation in Algebraic Combinatorics}

Basically, the architecture of a software for computing in algebraic
combinatorics is composed of two parts:
\begin{itemize}
\item a computer algebra kernel dealing with the bookkeeping of expressions
  and linear combinations (parsing, printing, collecting, Gaussian and
  Groebner elimination algorithm\dots);
\item a very large bunch of small combinatorial functions which enumerate and
  manipulate the combinatorial data structures.
\end{itemize}
In algebraic combinatorics software, for each basic combinatorial structure
such as permutations or partitions, there are typically 50-200 different
functions. Conjugating a partition is a very good example of what those
many functions do, that is surgery on lists of integers or lists of lists of
integers or more advanced recursive structures like trees\dots{} In a basic
computation, most of the time is spent mapping or iterating those functions
on some sets of objects. But due to combinatorial explosion those sets can be
very large so these functions must be very well optimized.

\subsection{Properties}
The definition of conjugate (diagonal symmetric of its partition shape)
is easy to understand but may conduct to naive implementations
that may be inefficient.

Let us suppose that we represent an integer partition by an integer array
starting from $1$. For example $\lambda=(3,2,1,1,1)$ gives $t[1]=3$,
$t[2]=2$,...  $t[l(\lambda)]=1$. Recall that $t[i]$ is non-increasing, that is
$t[i+1]\leq t[i]$.

One way to compute the conjugate is to count boxes: in our previous 
example the first column of $\lambda$ had 4 boxes, the second had 3 etc.
Therefore, to compute the number of boxes in a column $j$ we need to know
how many lines are longer than $j$. As a consequence, if $t_c$ is the
array representing the conjugate, the following formula gives the value of the
entries of the conjugate:
$$
t_c[j]=\left|\{i\,|\, 1\leq i\leq l(\lambda) \wedge t[i]\geq
  j\}\right|\,.
$$
Note that $t_c[j]=0$ if $j>t[1]$, so the previous expression must be computed
only from $j=1$ to $j=t[1]$.
This last property will be one of our predicates used to check
the correctness of loop invariants.

\subsection{SCHUR Implementation}
Here follows the code of the conjugate function extracted from the SCHUR
software. We expanded type definitions (C ``structs'' and ``typedefs'') 
from the original code just to simplify the work of Frama-C and
to make this part of code independent from the rest of the SCHUR software
(getting rid of global variables and so on).

\begin{center}
\begin{boxedverbatim}
#define MAX 100

void conjgte (int A[MAX], int B[MAX]) {
  int i, partc = 1, edge = 0;

  while (A[partc] != 0) {
     edge = A[partc];
     do 
          partc = partc + 1;
     while (A[partc] == edge);
     for (i = A[partc] + 1; i <= edge; i++)
          B[i] = partc - 1;
  }
}
\end{boxedverbatim}
\end{center}

Note that this implementation is not naive (and not so 
easy to understand) but its time complexity is optimal
(linear in the length of the partition). 

The algorithm is based on looking for 
the set of descents of the partition\footnote{
A descent is such that $t[i]<t[i-1]$}. The do--while
loop follows a ``flat'' portion of the partition ($t[i]=t[i-1]$)
until a descent is found. Next the for--loop assigns the values of the B
array 
according to the flat portion. The following figure clarifies
this: we have denoted {\tt partc}$_1$ the value of {\tt partc} at the entrance of
while loop. {\tt partc}$_2$ is the value of {\tt partc} after 
the do--while loop.
For clarity's sake we supposed {\tt A}[{\tt partc}$_2$]+1 to be different from 
{\tt A}[{\tt partc}$_1$].
If we count boxes column by column to construct array {\tt B}, it is clear
that {\tt B}[$i$]={\tt partc}$_2$-1 for all {\tt A}[{\tt partc}$_2$]+1 
$\leq i\leq$ {\tt A}[{\tt partc}$_1$]={\tt edge}.

\def\carre{$\DessinCarre{}$}
\def\pts{...}
\def\fbas{$\downarrow$}
\def\fdroit{$\rightarrow$}
\def\A{{\tt A}}
\def\partc{{\tt partc}}
\def\edge{{\tt edge}}
\begin{center}
\begin{tabular}{rcccc@{}ccccc}
              &       &\begin{sideways}1\end{sideways}&\pts&\begin{sideways}\A[\partc$_2$]\end{sideways}&\begin{sideways}\A[\partc$_2$]+1\end{sideways}&\pts&\begin{sideways}\A[\partc$_1$]\end{sideways}&\pts&\begin{sideways}\A[1]\end{sideways}\\
              &	      &\fbas &    &\fbas       &\fbas         &    &\fbas        &    &\fbas\\
1             &\fdroit&\carre&\pts&\carre      &\carre        &\pts&\carre       &\pts&\carre\\
\vdots        &	      &\vdots&    &\vdots      &\vdots        &    &\vdots\\
\partc$_1$&\fdroit&\carre&\pts&\carre      &\carre        &\pts&\carre\\
\vdots  &             &\vdots&    &\vdots      &\vdots        &    &\vdots\\
\partc$_1$+n   &\fdroit&\carre&\pts&\carre      &\carre        &\pts&\carre\\
\partc$_2$     &\fdroit&\carre&\pts&\carre\\
\vdots        &       &\vdots&    &\vdots\\
\end{tabular}
\end{center}

\section{The Formal Proof of the Conjugate Function}
\subsection{Annotations}\label{section:Annotations}

In the following paragraphs we present the annotations
added to the
code. Note that this is the only additions made to it.
First we have to specify the model of integers we want to deal
with:

\begin{verbatim}
#pragma JessieIntegerModel(strict)
\end{verbatim}

This means that int types are modeled by integers with appropriate bounds, 
and for each arithmetic operation, it is mandatory to show that no overflow occurs.

Next, we have to express in first-order logic what an
integer partition (stored in an array) is:

\begin{verbatim}
#define MAX 100
/*@ predicate is_partition{L}(int t[]) = 
    (\forall integer i; 1 <= i < MAX ==> 0 <= t[i] < (MAX-1)) &&
    (\forall integer i,j; 1 <= i <=j < MAX ==> t[j] <= t[i]) && 
    t[MAX-1]==0;
  */
\end{verbatim}

Note that annotations are coded in the C comments, starting with a \verb+@+.
The \verb+{L}+ term is the context (pre, post, etc.), 
we won't detail it here, see~\cite{Jessie,ACSL}
for details.

The data structure (array of integers) 
comes from the way the SCHUR software represents
integer partitions. 0 is used as a mark of end of array,
just like character strings in C. The MAX value
comes from the original source code as well.
The first line of the predicate \verb+is_partition+
expresses that we are able to compute the conjugate (if at least
one element is greater than or equal to MAX-1, the conjugate 
will no be able
to be stored in an array of size MAX-1 with the last element fixed to 0).
From the source code it is expressed by an external simple test
on \verb+t[1]+, but expressing it like that
simplifies automatic provers job. 
The second line of the predicate defines the non-increasing order.

The following predicate is needed to express how we count blocs to compute
the conjugate. It may be read as $z$ equals the number of elements of partition
$t$, whose indexes are included in $\{1,..,j-1\}$
and whose values are greater than or equal to $k$. It is theoretically
possible to express it as an axiomatic theory, a kind of function, but
automatic provers we use make a better use of predicates. Note that we need
to explicit the $z=0$ case, in order to be able to prove the global
post-condition \verb+is_conjugate(A,B)+.

\begin{verbatim}
/*@ predicate countIfSup{L}(int t[],integer j,integer k,integer z)=
    is_partition{L}(t) && 
    1<= j <= MAX &&
    1<= k < MAX &&
    ((1<=z<j && \forall integer i ;  1<=i<=z ==> t[i]>= k)
     || (z==0 && \forall integer i ; 1<=i<j ==> t[i]<k)) ;
 */
\end{verbatim}

Here is what we want to obtain at the end of the computation,
t2 is a conjugate of t1 if the following holds:

\begin{verbatim}
/*@ predicate is_conjugate{L}(int t1[], int t2[]) =
    \forall integer k ; 1<=k<MAX ==> countIfSup(t1,MAX,k,t2[k]);
 */
\end{verbatim}

Finally, here is the function. First we have
to give precise requirements on the inputs. For example,
(\verb|\valid(A+ (1..(MAX-1)))| means that memory has been allocated so array
indexes from 1 to MAX-1 are allowed).  From the original code, the B array is
supposed to be initialized with zeros before calling the function. This is
translated into a \verb+requires+ directive. Next, we specify which memory
elements are modified by the function (\verb+assigns+). This is used for
safety proofs. In the end, the output is correct if the post-condition
(\verb+ensures+) is met.

\begin{verbatim}
/*@ requires \valid(A+ (1..(MAX-1)));
    requires \valid(B+ (1..(MAX-1)));
    requires is_partition(A);
    requires \forall integer k; 1<=k<MAX ==> B[k]== 0;
    assigns B[1..A[1]];
    ensures is_conjugate(A,B);
    */
void conjgte (int A[MAX], int B[MAX])
{
  int i, partc=1, edge = 0 ;
\end{verbatim}

Now we have to define the loop variant and invariant for each loop
(to prove properties).
The ``loop variant'' must decrease, while remaining non negative, 
to be able to prove termination. 
We also use a ``\verb+ghost+ variable'' to store the state of a variable
before any modification.

\begin{verbatim}
  /*@ loop variant MAX-partc;
      loop invariant 1<=partc<MAX;
      loop assigns B[1..A[1]];
      loop invariant \forall integer k; 
      	  A[partc]+1 <=k <= A[1] ==> countIfSup(A,MAX,k,B[k]);
    */
  while (A[partc] != 0) {
     edge = A[partc];
   
     /*@ ghost int old_partc = partc; */

    /*@ loop variant MAX-partc;
        loop invariant old_partc<=partc ;
        loop invariant \forall integer k; 
            old_partc<= k <= partc ==> A[k]==edge;
        loop invariant partc<MAX-1;
     */
     do
         partc = partc + 1;
     while (A[partc] == edge);
\end{verbatim}

We also use the \verb+assert+ directive to have a verification point
of a property that may help automatic provers for the next properties
or global ones.

\begin{verbatim}
     /*@ assert countIfSup(A,partc,edge,partc-1);*/

     /*@ loop variant edge-i;
         loop invariant i >= A[partc]+1 && edge+1>=i ;
         loop invariant \forall integer k; 
             A[partc]+1 <=k <i ==> countIfSup(A,MAX,k,B[k]);
         loop assigns B[ (A[partc]+1)..edge];
       */
     for (i = A[partc] + 1; i <= edge; i++)
         B[i] = partc - 1;
  }
}
\end{verbatim}

\subsection{Proofs}\label{section:proof}

\begin{figure}[htb]
\centerline{\includegraphics[width=0.95\textwidth]{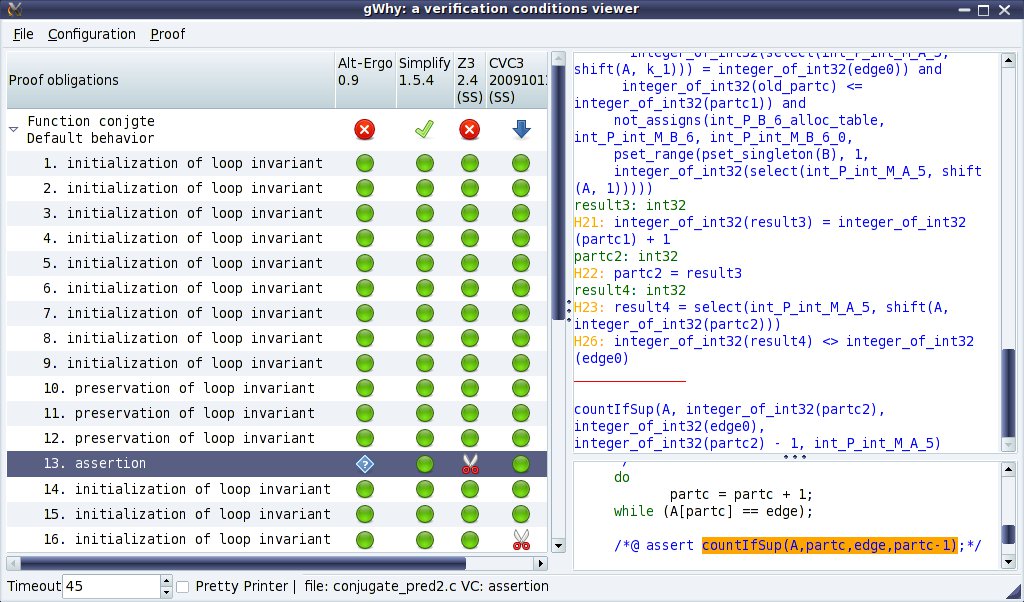}}
\caption{Graphical Interface: default behavior}\label{fig:snapshot1}
\end{figure}

The figures 1 to 3 are snapshots of gWhy (Frama-c graphical interface 
when using plugin Jessie). We applied this tool on the previous
annotated code.

The Verification Conditions (VC, also called proof obligations)
that have to be proved one by one
(line by line) appear to the left of 
each of the following snapshots.
In the upper right part of the window, we can check at a glance
what hypotheses 
are known and what is to be proved at the bottom of it
(under the line).
No circularity paradox is
possible here, since the proof of a VC can only rely on other 
VC higher in the control-flow graph of the function.

In the lower right part of the window, the corresponding part of
the annotation is highlighted in the source code 
with some lines before and after it.

\begin{figure}[htb]
\centerline{\includegraphics[width=0.95\textwidth]{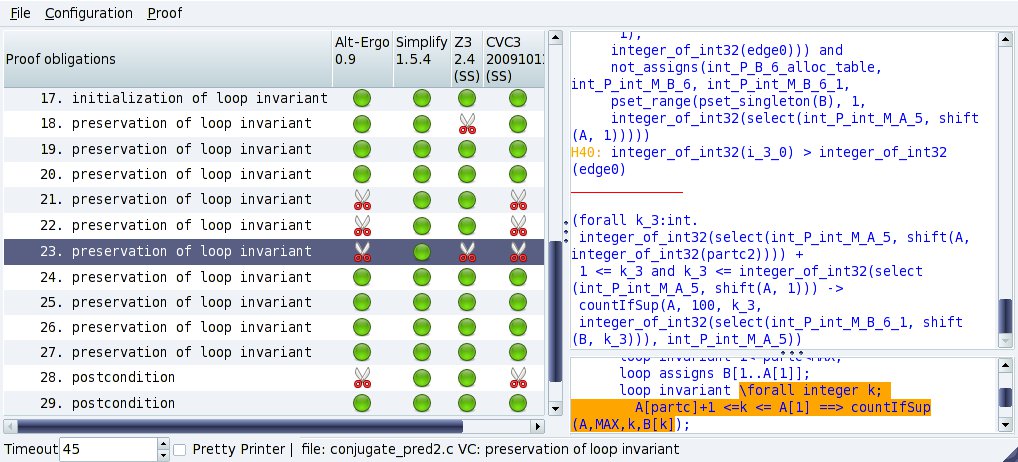}}
\caption{Graphical Interface continued}\label{fig:snapshot2}
\end{figure}

We will now focus on the VC part, to the left. We can see (green)
dots meaning that this property has been proved by this prover.
There is also (blue) rhombus with a question mark inside (see assertion 13),
indicating that this prover will not be able to to prove
this property. Actually, this does not mean that this VC is wrong, 
remember that these provers use heuristics. Sometimes, you may see scissors
meaning that the maximum execution time has been reached without proving
the VC. Again, this does not mean that the corresponding VC is
wrong. Finally, at the top of a column a (green) check or (red, with a white
cross inside) point is shown. The first one means that all properties have been
proved by that prover.  In fig.\ref{fig:snapshot1}, The (blue) arrow at the top of the
CVC3 column means
that it is still computing some unshown VC (greater than number 16).

The last figure is the final part. The provers have worked on
the safety of the code, that is to say, integer bounds (overflow problems),
pointer referencing and termination.

\begin{figure}[htb]
\centerline{\includegraphics[width=0.95\textwidth]{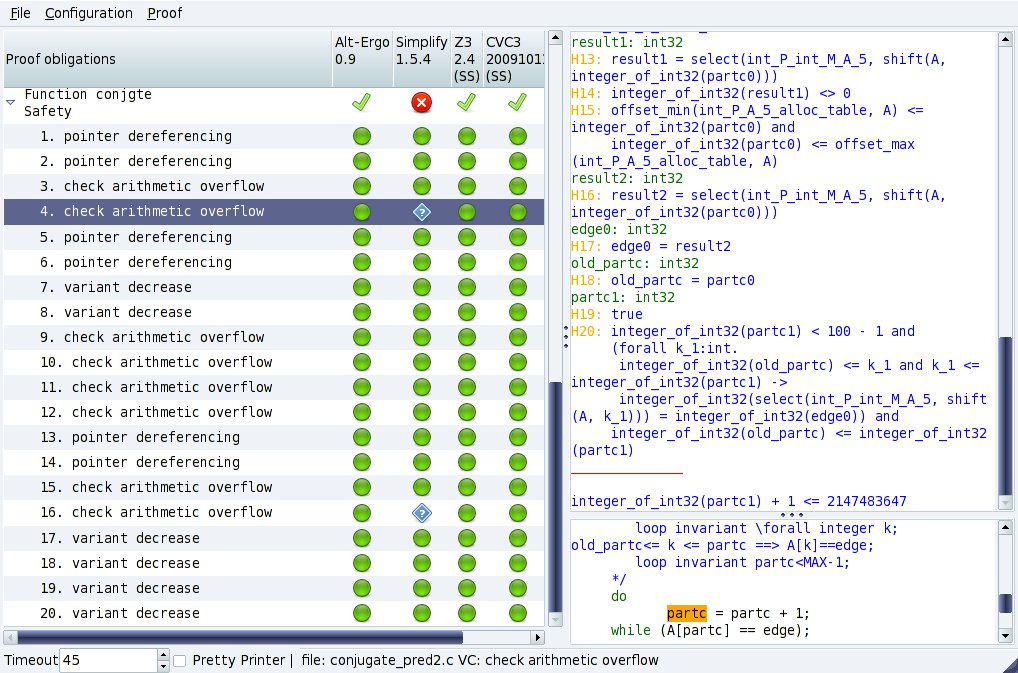}}
\caption{Graphical Interface: Safety}\label{fig:snapshot3}
\end{figure}

As seen in section~\ref{section:Annotations}, the \verb+B+ array has to be
initialized with zeros before calling the function. This requirement has been
enlightened thanks to the annotations and tools, in particular because without
the line \verb+requires \forall integer k;1<=k<MAX ==> B[k]==0+, the postcondition
which states that \verb+B+ is a conjugate of \verb+A+ cannot be proved.

We have also used Coq proof assistant. However, it not being essential
to our present point, we chose to live aside the detail
of this procedure (see section~\ref{section:Difficulties}).

\subsection{Problems, Mistakes}\label{section:Difficulties}

As usual when using formal proof tools, there are several ways to
formalize or to annotate programs. Choices made during at this stage are
very important for future proofs. For example, declaring a function as
an axiomatic theory or as a predicate will
suppose corresponding proofs to be different. 
We can make a similar remark with
data-types used in programs.

For these reasons, using ``good'' annotations
which allows automatic provers to prove verification conditions (VC)
successfully is a clever way to go about it.
 
When we deal with 40,000 lines of undocumented code, another critical
part of the work consists in ``correctly'' isolating the piece of code
that we want to prove. The code can use global variables, 
initializations made
by other functions, or use intricate data-types and so on.

In the following paragraphs, these problems
and associated mistakes are discussed.

\subsubsection{Isolating a Part of Program.}
Generally speaking, the analyzed function must be free of
external calls. More precisely if a function is called from it, 
it has to be incorporated in the code (like macro expansion)
or, at least, independently proved.

Next, data types must be simplified. Even if Frama-C can cope
with simple structures, it is better to have a first pass 
on them (unions suppression, typedef expansion and so on).

\subsubsection{How to Make Good Annotations?}

As previously explained, ACLS is a language which is used to annotate
C programs. Annotating an existing program consists in choosing
properties (comportment, results,...) that the user wants to be
``confirmed", such as preconditions, loop invariants, post
conditions. In our case, for example, one of the most relevant
properties we proved is that {\em the result B is the conjugate of
the partition A}. This property is stated as a postcondition. 

As usual, there are several ways to formalize annotations. 
Particularly when using external provers, a good method is
to know how provers work. Here, we have to remember that the
automatic provers are SMT solvers (see section~\ref{section:Frama-C}). 

As an example, we can give the definition of \verb+countIfSup+. In a first
formalization we wrote it as an ``axiomatization''. 
But due to another problem that we will describe
in the next paragraph, we needed to make some proofs in Coq which
used \verb+countIfSup+. Then, to make it easier for Coq, we decided to try to
define it inductively. Thanks to this other definition, some
conditions were automatically proved by SMT solvers. This
example shows how important formalization choices can be.

In the next paragraph, we will explain and illustrate
how Coq allowed us to correct some errors in our annotations.

\subsubsection{Why Coq?} 

Once annotations are completed, the method consists in using automatic
provers (using gWhy for example). As previously explained, if all
proof obligations are proved by at least one prover, the work can be
considered as finished. But, if one or more proof obligations is/are still
unproved, several approaches are possible: the first one consists in
verifying that annotations are ``sufficient'', that is to say a precondition or
a loop invariant is not missing. Another approach, when the user suppose
that his annotations
are correct, is to use an external non automatic
prover to try to prove proof obligations that have not been verified
previously. 

In our case, we used the interactive theorem prover Coq twice. The first time was because a
postcondition had not been proved by SMT provers. When we began Coq
proof, we discovered that the definition of \verb+countIfSup+ was
incomplete: the second part of the ``\verb+||+'' (logical or) was missing.

The second time we used Coq was to prove a loop
invariant. Similarly, we detected another incompleteness 
in \verb+countIfSup+
definition ($j < MAX$ instead of $j \leq MAX$). Proof assistants are
well adapted to detect this kind of problems. 
Indeed, building formal proofs manually, a user can
easily see which hypotheses are necessary.

After having corrected and replaced the ``axiomatization''
of \verb+countIfSup+ by a predicate, all proof obligations
have been proved by at least one automatic prover.

Note that the new definition allowed us to remove from the
annotations one
additional lemma which, at first, appeared necessary.

\subsubsection{Other Vicissitudes.}

Among the main encountered difficulties, we can mention the confidence
in the provers we used. In our case, one of the versions of CVC3 was faulty
and proved all VC correct, even when they were false. For this reason we
decided to consider that a proof obligation was proved when 
at least two automatic provers succeed on proving it. 
It is the case for all our obligations except one
(VC \# 23 is only proved by Simplify). 
The proof of VC \# 23 is in progress using Coq. 

\section{Conclusion and Future Work}

We have isolated and formally proved one of the
key commands of the SCHUR software.
This work reinforced us in the idea
of formally proving chosen parts of software of the same kind,
composed of 40,000 lines of undocumented code.

Thanks to this approach, we have focused on critical points (such as
particular initializations of arrays and appropriate bounds) from the
original code and by extension, 
we have understood the progression axis of the
methodology. In particular, it is better to know how SMT automatic
provers work to try to make a ``good'' annotation so that obligation
proofs will be more easily proved by them.  In the methodology, non
automatic external provers like Coq may be used to refine annotations,
and to prove obligations when no automatic provers succeed.

The conjugate function is a basic brick of combinatorics.  This give us
perspective to prove other functions.  
Therefore, as a future work, the second
step is to prove algorithms relying on exhaustive enumeration algorithm,
 such
as computation of Littlewood-Richardson coefficients, Koskas numbers, Koskas
matrices, representation multiplicity in tensor product decompositions, etc.

The final objective will be to build proved libraries usable for
scientific community.

\bibliography{biblio}
\bibliographystyle{splncs}

\end{document}